\documentstyle[12pt]{article}

\newcommand{\be}{\begin{equation}}
\newcommand{\ee}{\end{equation}}
\newcommand{\bn}{\begin{eqnarray}}
\newcommand{\en}{\end{eqnarray}}

\normalbaselines
\begin{document}
\begin{titlepage}
\begin{flushright}
{\bf HU-SEFT \ R \ 1995-13}
\end{flushright}
\begin{center}

\vspace*{1.0cm}

{\Large{\bf Evolution of QCD Coupling Constant at Finite Temperature in the
 Background Field Method}}

\vskip 1.5cm

{\large {\bf M. Chaichian}}

\vskip 0.5cm

High Energy Physics Laboratory, Department of Physics \\
and Research Institute for High Energy Physics, \\
University of Helsinki\\
P.O.Box 9 (Siltavuorenpenger 20 C), FIN-00014  \\
Helsinki, Finland\\

\vskip 0.2cm

$and$

\vskip 0.2cm

{\large {\bf M.Hayashi}}

\vskip 0.5cm

School of Life Science, Tokyo University of Pharmacy and Life Science, 1432-1
Horinouchi, Hachioji, Tokyo 192-03, Japan

\end{center}

\vspace{3 cm}

\begin{abstract}
\normalsize

The evolution of QCD coupling constant at finite temperature is considered by
making use of the finite temperature renormalization group equation up to the
one-loop order in the background field method with the Feynman gauge and the
imaginary time formalism. The results are compared with the ones obtained in
the literature. We point out, in particular, the origin of the discrepancies
between different calculations, such as the choice of gauge, the break-down of
Lorentz invariance, imaginary versus real time formalism and the applicability
of the Ward identities at finite temperature.

\end{abstract}
\end{titlepage}

\renewcommand{\theequation}{\arabic{equation}}
\section{Introduction}
\setcounter{equation}{0}
\renewcommand{\theequation}{1.\arabic{equation}}

One of the application of field theory at finite temperature [1-4] is to find
the behaviour of coupling constant as a function of energy, temperature, and
the chemical potential using the renormalization group (RG) equation. The
knowledge of coupling constant at finite temperature environment then can be
used, for instance, in perturbative calculations for quark-gluon plasma
created in ion-ion collisions at high energies, in the evaluation of the grand
unification scale in a cosmological context and the shift of the energy levels
in a hydrogen plasma.

Recently a great deal of attention has been paid to the study of the behaviour
 of the coupling contant at finite $T$ [5-19]. The resulting formula for
 the temperature and scale dependent part of the coupling constant has proved
 to be very sensitive to the prescription chosen. In this paper we wish to
 examine this and clarify the origin of the discrepancies between different
 calculations in the literature. We employ the background field method [20,21],
which is based on a manifestly gauge invariant generating functional, and the
Feynman gauge. The background field method provides notorious simplifications
since we have to calculate only the renormalization constant of the background
field gluon propagator. We discuss this method in section 2.1. Since the
coupling constant depends not only on energy but also on temperature, we derive
a pair of RG equations - one for energy and one for temperature. This is done
in section 2.2 and in section 2.3 we derive its solution. In section 2.4 we
discuss the Feynman rules at finite temperature in the imaginary time
formalism.

In order to get the renormalization constant of the background field gluon
propagator we have to calculate the polarization tensor. At finite temperature
due to the lack of the Lorentz invariance the structure of the polarization
tensor is not equivalent to the zero temperature one. In section 3.1 we give
a prescription how to define the renormalization constant at finite
temperature. In section 3.2 we calculate the polarization tensor in the
one-loop approximation. In section 3.3 calculating the vacuum part of the
polarization tensor we reproduce the well known zero temperature formula, i.e.
the standard vacuum QCD one-loop result [22,23], which reads as

\begin{eqnarray}
[g^{2}(\mu)]^{-1}=[g_{0}^{2}(\mu_{0})]^{-1}[1+2g_{0}^{2}(\mu_{0})K_{0}/
(4\pi)^{2}]\ ,
\end{eqnarray}
where
\begin{equation}
K_{0}=(11N/3-2n_{F}/3)\ln(\mu/\mu_{0}).
\end{equation}
Here $N$ is for $SU(N)$, $n_{F}$ is the number of flavours, $\mu$ is the energy
scale and $\mu_0$ is the reference scale.

We calculate the matter part of the polarization tensor in section 4.1 and, in
section 4.2, we derive the temperature and scale dependent part of the
coupling constant.

In section 5.1 we review the results of the previous works and in 5.2 we
compare the asymptotic expansion formulas at $a=\mu/T\ll 1$ derived in
different schemes. We give a few comments and, in particular, discuss the
origin of the discrepancies between different calculations.

In Appendix we collect the relevant integrals which we encounter in the text.

\vspace{5mm}
\section{Formal methods}
\subsection{Background field method}
\setcounter{equation}{0}
\renewcommand{\theequation}{2.\arabic{equation}}

Explicit gauge invariance, which is present at the classical level in gauge
field theories, is normally lost at the quantum level. This can be seen from
the generating functional for QCD:
\begin{eqnarray}
&&Z(J)=\int DA D\Psi D\bar{\Psi} D\eta D\bar{\eta}\exp\{i\int d^{4}x
[{\cal L}_{QCD}(A,\Psi,\bar{\Psi})\nonumber\\
&&-(G^{a})^{2}/(2\xi)+J^{\mu a}A_{\mu a}+\bar{\eta} \partial G^{a}/\partial
\omega^a\eta]\},
\end{eqnarray}
where $A$ is the gluon field, $\Psi$ and $\bar{\Psi}$ are the fermion fields,
$\eta$ and $\bar{\eta}$ are ghost fields, $(G^a)^2$ is a gauge fixing term
($\xi$:gauge parameter), $\omega^a$ is a $SU(3)$ group parameter, and
$J^{\mu a}A_{\mu a}$ is a source term. In $SU(3)$ the gauge fields transform
in the following way:
\begin{equation}
A_{\mu} \rightarrow U(A_{\mu}-i\partial_{\mu}/g)U^{\dagger},
\end{equation}
\begin{equation}
\Psi \rightarrow U\Psi, \bar{\Psi} \rightarrow \bar{\Psi} U^{\dagger},
\end{equation}
\begin{equation}
\eta \rightarrow U\eta, \bar{\eta} \rightarrow \bar{\eta} U^{\dagger},
\end{equation}
where $U(x)=\exp [i\omega(x)^{a}T_{a}]$ is a unitary transformation, and
$T_{a}$ is a generator for $SU(3)$. By using transformations (2.2)-(2.4) one
can see that the gauge invariance of $Z(J)$ is lost in commonly used gauges
such as $\partial_\mu A_\mu=0$.

The advantage of the background field method [20,21] is that it can maintain
the explicit gauge invariance. For this purpose we devide the gluon field
$A_\mu$ into a sum of a classical background field $B_\mu$ and a quantum field
$Q_\mu$
\begin{equation}
A_{\mu}=B_{\mu}+Q_{\mu},
\end{equation}
and choose for the gauge fixing term $G^{a}$ the background field gauge
condition
\begin{equation}
G^{a}=(D_{\mu}(B))^{ab}Q^{\mu}_{b},
\end{equation}
where $D_{\mu}$ is the covariant derivative:
\begin{equation}
(D_{\mu})_{ab}=\partial_{\mu}\delta_{ab} +gf_{abc} B_{\mu}^{c}.
\end{equation}
The generating functional reads
\begin{eqnarray}
&&Z(J,B)=\int DQ D\Psi D\bar{\Psi} D\eta D\bar{\eta} \int \exp\{i\int d^{4}
x[{\cal L}_{QCD}(B+Q,\Psi,\bar{\Psi})\nonumber\\
&&-(G^{a})^{2}/(2\xi)+J_{\mu}^{a}A_{\mu}^{a}+\bar{\eta}\partial G^{a}/\partial
\omega^a\eta]\}.
\end{eqnarray}
This functional is gauge invariant, which follows from the transformations:
\begin{equation}
B_{\mu} \rightarrow U(B_{\mu}-i\partial_{\mu}/g)U^{\dagger},
\end{equation}
\begin{equation}
Q_{\mu} \rightarrow UQ_{\mu}U^{\dagger},
\end{equation}
\begin{equation}
D_{\mu}(B) \rightarrow UD_{\mu}(B)U^{\dagger},
\end{equation}
\begin{equation}
J_{\mu}(B) \rightarrow UJ_{\mu}(B)U^{\dagger},
\end{equation}
together with Eqs.\hspace{1mm}(2.3) and (2.4). Notice that we are only
changing the integration variables in Eq. (2.8).

In the background field method the quantum gauge fields and the ghost field
need not be renormalized since they appear only inside loops. No vertex
functions are to be considered. Thus only renormalizations
\begin{equation}
g_{0}=Z_{g}g_{R},
\end{equation}
\begin{equation}
B_{0}=(Z_{B})^{1/2}B_{R},
\end{equation}
\begin{equation}
\xi_{0} =Z_{\xi}\xi_{R},
\end{equation}
are needed.

The explicit gauge invariance of $Z(J,B)$ implies that the perturbation series
 is gauge invariant in every order in $g_R$ and that the background field
renormalization factor $Z_B$ and the coupling constant renormalization $Z_g$
are related with each other. The field tensor $F_{\mu\nu}$ in ${\cal L}_{QCD}$
has to be gauge covariant and is renormalized as
\begin{eqnarray}
F_{\mu \nu}^{a} =(Z_{B})^{1/2}[\partial_{\mu}(B_{\nu}^{a})_{R}-\partial_{\nu}
(B^a_\mu)_R+g_RZ_g(Z_B)^{1/2}f^{abc}(B^b_{\mu})_R(B^c_{\mu})_R].
\end{eqnarray}
Thus we have a relation
\begin{equation}
Z_{g}=(Z_{B})^{-1/2},
\end{equation}
which enables us to calculate the evolution of coupling constant $g_{R}$ in a
simple way. We assume this relation to be valid also at finite temperature.
\vspace{5mm}

\subsection{Derivation of the RG equations}

{}From Eqs.\hspace{1mm}(2.13) and (2.17) we have
\begin{equation}
g_{0}=g_{R}(Z_{B})^{-1/2}.
\end{equation}
We use the dimensional regularization [24,23] and perform our calculations in
$(4-2\varepsilon)$ dimensions. We notice from the Lagrangian of QCD that the
dimension of
a gluon (quark) field is $\mu^{1-\varepsilon}(\mu^{3/2-\varepsilon})$, where
$\mu$ is the scale parameter. We redefine $g_0$ so that it becomes
dimensionless
and rewrite Eq. (2.18) as
\begin{equation}
g_{0}=g_{R}(Z_{B})^{-1/2}\mu^{\varepsilon},
\end{equation}
where the bare coupling $g_{0}$ does not depend on temperature $T$ and the
scale $\mu$. A pair of RG equations results upon taking the derivative of
Eq. (2.19) with respect to $T$ and $\mu$
\begin{equation}
T\partial/\partial T[g_{R}Z_{B}^{-1/2}\mu^{\varepsilon}]=0,
\end{equation}
\begin{equation}
\mu\partial/\partial \mu[g_{R}Z_{B}^{-1/2}\mu^{\varepsilon}]=0.
\end{equation}
These equations, which are not symmetrical in $T$ and $\mu$, determine the
behaviour of coupling constant $g_R$ with respect to $T$ and $\mu$ changes.
Generally $Z_B$ has the form [22,23]
\begin{equation}
Z_{B}=1+\sum_{i=1}^{\infty}g_{R}^{2i}\ [\sum_{j=1}^{i}
A_{i}^{(j)}/\varepsilon^j+f^{(i)}(\mu,T)],
\end{equation}
where $A_{j}^{(j)}/\varepsilon^{j}$ are the divergent contributions independent
on $T$ and $\mu$, while $f^{(j)}(\mu,T)$ are convergent temperature and scale
dependent functions which vanish in the low-temperature limit.

{}From Eqs.\hspace{1mm}(2.21) and (2.22) one can derive the following formula,
by
taking $\varepsilon\rightarrow 0$, (from now on we write $g=g_R$, for
 simplicity)
\begin{eqnarray}
\mu\partial g/\partial\mu=g^{3}\{-A^{(1)}+(\mu/2)\partial/\partial\mu[f^{(1)}
(\mu,T)]\}+O(g^5).
\end{eqnarray}
This equation is reduced to the ordinary one-loop RG equation in the zero-
temperature limit since in this limit the function $f^{(1)}(\mu,T=0)$ vanishes.

Similarly one can derive the RG equation for $T$ from Eqs.\hspace{1mm}(2.20)
and (2.22)
\begin{eqnarray}
T\partial g/\partial T=(g^{3}/2)T\partial/\partial T[f^{(1)}(\mu,T)]+O(g^{5}).
\end{eqnarray}
Equations (2.23) and (2.24) constitute the coupled RG equations.
\vspace{5mm}

\subsection{Solution of the coupled RG equations}

The solution to Eqs.\hspace{1mm}(2.23) and (2.24) can be found in a
straightforward way by integrating Eq. (2.23) from $\mu_0$ to $\mu$ and Eq.
(2.24) from $T_0$ to $T$. We have
\begin{eqnarray}
&&1/[2g^{2}(\mu_{0},T)]-1/[2g^{2}(\mu,T)]=-A^{(1)}\ln(\mu/\mu_{0})\nonumber \\
&&+[f^{(1)}(\mu,T)-f^{(1)}(\mu_{0},T)]/2,
\end{eqnarray}
\begin{eqnarray}
1/[2g^{2}(\mu,T_{0})]-1/[2g^{2}(\mu,T)]=[f^{(1)}(\mu,T)-f^{(1)}(\mu,T_{0})]/2.
\end{eqnarray}
Putting $T$ equal to $T_{0}$ in Eq.\hspace{1mm}(2.25), and using
Eq.\hspace{1mm}(2.26) we arrive at the desired solution
\begin{eqnarray}
&[g^{2}(\mu,T)]^{-1}=[g^{2}(\mu_{0},T_{0})]^{-1}+2A^{(1)}\ln(\mu/\mu_{0})
\nonumber\\
&-[f^{(1)}(\mu,T)-f^{(1)}(\mu_{0},T_{0})].
\end{eqnarray}
This equation describes the evolution of the QCD coupling constant as a
function of the momentum scale and the arbitrary temperature [5-9].
 $(\mu_0,T_0)$ denotes the reference point.
\vspace{5mm}

\subsection{Imaginary time formalism}

The Feynman rules at finite temperature $T(=1/\beta)$ are derived from the
 generating functional
\begin{eqnarray}
&&Z(J)=c \int DQ D\Psi D\bar{\Psi} D\eta D\bar{\eta} \exp[\int_{0}^{\beta}
d\tau\int d^3x\{{\cal L}(x,\tau)\nonumber\\
&&-[G^{a}(x,\tau)]^{2}/(2\xi)+\bar{\eta}(x,\tau)\partial G^{a}/\partial
\omega^{a}\eta(x,\tau)\}],
\end{eqnarray}
where $\eta,\bar{\eta},Q$ are periodic fields, while $\Psi,\bar{\Psi}$ are
antiperiodic fields.

This functional differs from Eq.\hspace{1mm}(2.8) only in that time $\tau$ is
now imaginary {\it it}. Infinite time domain has been compactified to the
finite interval [$0,\beta$]. From Eq. (2.28) one can formulate the Feynman
rules at finite temperature by modifying the ones at zero temperature in the
following way [25]: For the loop integral we have
\begin{eqnarray}
\int d^{4}p/(2\pi)^{4}  \rightarrow (i/\beta)\sum_{n}\int d^{3}p/(2\pi)^{3},
\end{eqnarray}
and for the time-component of 4-momentum we have for bosons and ghosts:
\begin{eqnarray}
p_{0} =i2n\pi/\beta,
\end{eqnarray}
and for fermions:
\begin{eqnarray}
p_{0} =i(2n+1)\pi/\beta+\mu_{ch},
\end{eqnarray}
where $\mu_{ch}$ is a chemical potential. The frequency sums for bosons and
fermions in Eq. (2.29) are readily changed to contour integrals [26]: We have
for bosons and ghosts
\begin{eqnarray}
&&(i/\beta)\sum_{n=-\infty}^{\infty} f(p_{0}=i2n\pi/\beta)=(1/2\pi)
\int_{-i\infty+\varepsilon}^{i\infty+\varepsilon}dp_of(p_o)N_B(p_0)\nonumber\\
&&-(1/2\pi) \int_{-i\infty-\varepsilon}^{i\infty-\varepsilon} dp_{0}f(p_{0})
N_{B}(-p_0)+(1/2\pi)\int^{i\infty}_{-i\infty}dp_0f(p_0),
\end{eqnarray}
where
\begin{equation}
N_{B}(p_{0})=1/[\exp(\beta p_{0})-1],
\end{equation}
and for fermions
\begin{eqnarray}
&&(i/\beta)\sum_{n=-\infty}^{\infty} f[p_{0}=i(2n+1)\pi/\beta+\mu_{ch}]=
-(1/2\pi)\nonumber\\
&&\times\int_{-i\infty+\varepsilon}^{i\infty+\varepsilon} dp_{0}f(p_{0}+
\mu_{ch})N_{F}(p_0)+(1/2\pi)\int^{i\infty-\varepsilon}_{-i\infty-\varepsilon}
dp_0\nonumber\\
&&\times f(p_{0}+\mu_{ch})N_{F}(-p_{0})+(1/2\pi) \int_{-i\infty}^{i\infty}
dp_{0}f(p_0),
\end{eqnarray}
where
\begin{equation}
N_{F}(p_{0})=1/[\exp(\beta p_{0})+1].
\end{equation}
The first two terms in the right-hand side of Eq.\hspace{1mm}(2.34) correspond
 to particle and anti-particle contributions respectively and vanish at $T=0$.
 Since we will be only interested in phenomena in an enviroment where the sum
 of quantum numbers is not conserved, hereafter we put $\mu_{ch}$ equals to
 zero. Also we have omitted the term which gives the finite density
 contribution at $T=0$.

\vspace{5mm}
\section{Polarization tensor}
\subsection{Structure of the polarization tensor}
\setcounter{equation}{0}
\renewcommand{\theequation}{3.\arabic{equation}}

At a zero temperature environment the polarization tensor is Lorentz invariant
and can be expressed as [22]
\begin{equation}
\Pi^{ab}_{\mu \nu}=\delta^{ab}\Pi_{\mu \nu},
\end{equation}
where
\begin{equation}
\Pi_{\mu \nu}= \Pi(k^{2}g_{\mu \nu}-k_{\mu}k_{\nu}).
\end{equation}
The zero temperature polarization tensor is transverse with respect to $k$
(current conservation):
\begin{equation}
k_{\mu}\Pi_{\mu \nu}=0.
\end{equation}
At finite temperature, in the presence of matter, the Lorentz invariance is
lost and the polarization tensor can only be $O(3)$ rotational invariant
[27,17]. Then the polarization tensor can generally depend only on 4
independent
quantities, which we can choose, for example, $\Pi_{00}, k_i\Pi_{0i}$ and the
two scalars $\Pi_L$ and $\Pi_T$ appearing in
\begin{eqnarray}
\Pi_{ij}= \Pi_{T}(\delta_{ij}-k_{i}k_{j}/{\bf k}^{2})+ \Pi_{L}k_{i}k_{j}/
{\bf k}^2.
\end{eqnarray}
At finite temperature whether the transversality condition is satisfied or not
depends on the gauge used. The polarization tensor is not transversal, e.g.,
in the Coulomb gauge $(\partial_iA_i=0)$, but is transversal at the one-loop
level in the temporal axial gauge $(A_0=0)$, and in every order of the
perturbative calculations in the background field gauge [17]. The
transversality
condition (3.3) restricts the structure of the polarization tensor. From Eq.
(3.3) we have
\begin{equation}
k_{0}\Pi_{00}=k_{i}\Pi_{i0},
\end{equation}
and
\begin{equation}
k_{0}^{2}\Pi_{00}=k_{i}k_{j}\Pi_{ij}.
\end{equation}
Using Eqs.\hspace{1mm}(3.4) and (3.6) we obtain
\begin{equation}
\Pi_{L}=k_{0}^{2}\Pi_{00}/{\bf k}^{2}.
\end{equation}
For the coefficient of the transversal part of the polarization tensor we get
from Eqs. (3.4) and (3.7) an expression
\begin{equation}
\Pi_{T}=(\Pi_{ii}-k_{0}^{2}\Pi_{00}/{\bf k}^{2})/2.
\end{equation}
By using $\Pi_{ii}=(\Pi_{\mu \mu}-\Pi_{00})$, we rewrite Eq.\hspace{1mm}(3.8)
as
\begin{eqnarray}
\Pi_{T}=[\Pi_{\mu \mu}-\Pi_{00}(1+k_{0}^{2}/{\bf k}^{2})]/2.
\end{eqnarray}
Thus in Eqs.\hspace{1mm}(3.4)-(3.9) we have derived the general form of the
temperature dependent $O(3)$ symmetric polarization tensor in the background
field method.

The polarization tensor can be splitted into a sum of a temperature independent
(vacuum) part and a temperature and scale dependent (matter) part [4]:
\begin{equation}
\Pi_{\mu \nu}=\Pi_{\mu \nu}|_{vac}+\Pi_{\mu \nu}|_{matt} .
\end{equation}
At zero temperature limit the temperature dependent matter part vanishes.

The structure of the polarization tensor at zero temperature was given in
Eqs. (3.1) and (3.2). The polarization tensor is related to the gluon
propagator, $D_{\mu\nu}$, as
\begin{equation}
\Pi_{\mu \nu}=D_{R\mu \nu}^{-1}-D_{0 \mu \nu}^{-1}=\Pi D_{R\mu \nu}^{-1}.
\end{equation}
The relation between the bare and the renormalized propagators
(see Eq.\hspace{1mm}(2.14)) is
\begin{equation}
D_{0 \mu \nu}=Z_{B}^{-1}D_{R \mu \nu},
\end{equation}
which together with Eq.\hspace{1mm}(3.11) leads us to
\begin{equation}
Z_{B}=1-\Pi.
\end{equation}
Thus the renormalization constant $Z_{B}$ can be obtained from the (background
field) gluon self-energy tensor.

In investigating the behaviour of the coupling constant one has to include the
temperature dependent parts of the polarization tensor in the renormalization
constant $Z_B$ via Eq. (3.13). Although the finite renormalization is somehow
arbitrary one has to follow some rule or prescription in every order of
perturbation to be consistent [23].

At finite temperature we encounter another type of ambiguity which is caused
by the lack of the Lorentz invariance. In order to define the polarization
tensor at finite temperature we generalize Eq.(3.13) for $T=0$ as:
\begin{equation}
Z_{B}=1-\Pi|_{vac}-\Pi|_{matt},
\end{equation}
where we have either
\begin{equation}
\Pi|_{matt}=\Pi_{00}/{\bf k}^{2},
\end{equation}
or
\begin{equation}
\Pi|_{matt}=\Pi_{T}/{\bf k}^{2}.
\end{equation}
Naturally at $T=0$, Eq.(3.14) reduces to Eq.(3.13). $\Pi_{00}$ and $\Pi_{T}$
 are not connected with each other and in general there is no a priori way to
 decide which one is more natural.
\vspace{5mm}

\subsection{The polarization tensor at the one-loop level}

Our working formulas, which allow to analyze the evolution of QCD coupling
constant at finite temperature, are:Eq.(2.22), Eq.(2.27) and Eqs.\hspace{1mm}
(3.14)-(3.16) for $T\neq 0$ case [Eq.(3.13) for $T=0$ case]. Accordingly we
have
 to evaluate the self-energy diagrams in Fig.1 to get the renormalization
 factor for $Z_B$ up to the one-loop order.

The Feynman rules for the interaction vertices are the same as in the zero
temperature case, and therefore identical to those given in [21]. Evaluating
the one-loop diagrams (1a)-(1d) in the Feynman gauge (i.e., we set $\xi=1$ in
the gluon propagator) we find for the boson contributions in the polarization
tensor
\begin{eqnarray}
&&\Pi_{\mu \nu}|_{boson}=ig^{2}N\int d^{4}p/(2\pi)^{4}[4g_{\mu \nu}k^{2}+2
(k_\mu p_\nu+k_\nu p_\mu)\nonumber\\
&&+4p_{\mu}p_{\nu}-3k_{\mu}k_{\nu}-2(k+p)^{2}g_{\mu \nu}]/D,
\end{eqnarray}
where
\begin{equation}
D=(k+p)^{2}p^{2}.
\end{equation}
For the fermion loop, neglecting the quark masses compared to momentum scale
and temperature, we have from the diagram (1e):
\begin{eqnarray}
&&\Pi_{\mu \nu}|_{fermion}=-4ig^{2}T_{F}n_{F}\int d^{4}p/(2\pi)^{4}\{k_{\mu}
p_{\nu}\nonumber\\
&&+k_{\nu}p_{\mu}+2p_{\mu}p_{\nu}-g_{\mu \nu}[(kp)+p^{2} ]\}/D,
\end{eqnarray}
where $T_{F}=1/2$ for $SU(3)$. We have also included $n_{F}$ (number of
flavours) in Eq. (3.19).

Our polarization tensor satisfies the transversality condition
\begin{equation}
k_{\mu}\Pi_{\mu \nu}|_{boson}=k_{\mu}\Pi_{\mu \nu}|_{fermion}=0.
\end{equation}
This can be shown explicitly as follows: From Eq.\hspace{1mm}(3.17) we have
\begin{eqnarray}
&&\Pi_{\mu \nu}|_{boson}=ig^{2}N\int
d^{4}p/(2\pi)^{4}[-k_{\nu}/p^{2}\nonumber\\
&&-(k+p)_{\nu}/(k+p)^{2}+2p_{\nu}/p^{2}-p_{\nu}/(k+p)^{2}].
\end{eqnarray}
Replacing $k+p \rightarrow p$ in the second term we have
\begin{eqnarray}
\Pi_{\mu \nu}|_{boson}=-ig^{2}N\int d^{4}p/(2\pi)^{4}[(k-p)_{\nu}/p^{2}+
p_{\nu}/(k+p)^2].
\end{eqnarray}
Then by changing $k-p \rightarrow -p$ in the first term, we immediately come to
Eq.\hspace{1mm}(3.20) for the boson contributions. Similarly one can prove the
transversality condition for the fermion polarization tensor. In the next
subsection we will extract from $\Pi_{\mu \nu}$ the information on the vacuum
part of the polarization tensor.
\vspace{5mm}

\subsection{Vacuum part of the polarization tensor}

Using a standard technique and introducing the Feynman parametrization in
Eqs. (3.17) and (3.19), namely
\begin{equation}
1/(ab)=\int_{0}^{1} dx\{1/[ax+b(1-x)]^{2}\},
\end{equation}
we have
\begin{eqnarray}
1/D=1/[p^{2}(k+p)^{2}]=\int_{0}^{1} dx\{1/[(p+kx)^{2}+k^{2}x(1-x)]^{2}\}.
\end{eqnarray}
Changing variable $p \rightarrow p-kx$ we get the following formula for the
boson contributions from Eq. (3.17):
\begin{eqnarray}
&&\Pi_{\mu \nu}|_{boson}=ig^{2}N\int d^{4}p/(2\pi)^{4}\int_{0}^{1}
dx[4g_{\mu \nu}k^2+2(1-2x)(k_\mu p_\nu+k_\nu p_\mu)\nonumber\\
&&+4p_{\mu}p_{\nu}+(-3-4x+4x^{2})k_{\mu}k_{\nu}]/(p^{2}+K^{2})^{2},
\end{eqnarray}
where
\begin{equation}
K^{2}=k^{2}x(1-x).
\end{equation}
Similarly we get the formula for the fermion contributions from Eq.\hspace{1mm}
(3.19):
\begin{eqnarray}
&&\Pi_{\mu \nu}|_{fermion}=-2ig^{2}n_{F}\int d^{4}p/(2\pi)^{4}\int_{0}^{1}
dx[2(-x+x^2)k_\mu k_\nu+(1-2x)\nonumber\\
&&\times(k_{\mu}p_{\nu}+k_{\nu}p_{\mu})-g_{\mu \nu}(kp)+2p_{\mu}p_{\nu}+
g_{\mu \nu}k^2x]/(p^2+K^2)^2.
\end{eqnarray}
In these equations for the vacuum parts we have dropped the terms proportional
to $1/p^2$ and $1/(k+p)^2$, which turn out to be zero after simple calculation
[see Eq. (A.1) in the Appendix]. Notice that all of the integrals are
ultraviolet divergent and thus have to be regularized. For this purpose we
employ the dimensional regularization [24,23], which preserves gauge
symmetries explicitly. The integrals for the vacuum parts become Euclidean if
we change $ip_0\rightarrow  p_4$ and thus can be easily evaluated. We obtain
the following results:
\begin{eqnarray}
\Pi_{\mu \nu}|_{boson}=-11g^{2}N/(3\alpha)(g_{\mu \nu}k^{2}-k_{\mu}k_{\nu})/
\varepsilon +O(1),
\end{eqnarray}
and
\begin{eqnarray}
\Pi_{\mu \nu}|_{fermion} =2g^{2}n_{F}/(3\alpha)(g_{\mu \nu}k^{2}-k_{\mu}
k_{\nu})/\varepsilon +O(1),
\end{eqnarray}
where
\begin{equation}
\alpha=(4\pi)^{2}.
\end{equation}
Combining Eqs.\hspace{1mm}(3.28) and (3.29) and taking into account Eqs.(2.22),
(3.2) and (3.13), we get for $A^{(1)}$ in Eq. (2.27) as:
\begin{equation}
A^{(1)}=(11N-2n_{F})/(3\alpha),
\end{equation}
which is, as expected, in accord with Eq.\hspace{1mm}(1.2) for zero
temperature.
In concluding this subsection let us emphasize the simplicity of the
calculation by the background field method in contrast to the conventional
methods, e.g., in the covariant gauge [7-9].

\vspace{5mm}
\section{Temperature dependent parts of the polarization tensor}
\subsection{Calculation of $\Pi_{00}$ and $\Pi{\mu \mu}$}
\setcounter{equation}{0}
\renewcommand{\theequation}{4.\arabic{equation}}

In this subsection we calculate $\Pi_{00}$ and $\Pi_{\mu \mu}$, without using
the Feynman parametrization. To specify the subtraction point we employ the
static limit of zero external energy, which is commonly used in the literature
[5]. In this prescription the momentum $k$ is specified to be space-like
$k=(0,k_i)(i=1,2,3)$ with $k^2=-\mu^2$. Such a choice enables us to determine
the static properties. Using Eqs. (3.17) and (3.19) and the integrals from the
Appendix we obtain for $\Pi_{00}$
\begin{eqnarray}
&&\Pi_{00}=2ig^{2}N\int d^{4}p/(2\pi)^{4}[2k^{2}+2p_{0}^{2}-(p+k)^{2}]/D
|_{boson}\nonumber\\
&&+2ig^{2}n_{F}\int d^{4}p/(2\pi)^{4}[-2p_{0}^{2}+(kp)+p^{2}]/D
|_{fermion}\nonumber\\
&&=g^{2}T^{2}(N/6+n_{F}/12)-2g^{2}N\mu^{2}[4F_{B0}(a)-F_{B2}(a)]\nonumber \\
&&-2g^{2}n_{F}\mu^{2}[F_{F0}(a)-F_{F2}(a)],
\end{eqnarray}
with $a=\beta\mu$. Similarly we derive an expression for $\Pi_{\mu \mu}$:
\begin{eqnarray}
&&\Pi_{\mu \mu}=ig^{2}N\int d^{4}p/(2\pi)^{4}[5k^{2}-12(kp)-4p^{2}]/D|_{boson}
\nonumber\\
&&+4ig^{2}n_{F}\int d^{4}p/(2\pi)^{4}[(kp)+p^{2}]/D|_{fermion}\nonumber \\
&&=g^{2}T^{2}(N/3+n_{F}/6)-2g^{2}\mu^{2}[11NF_{B0}(a)+2n_{F}F_{F0}(a)],
\end{eqnarray}
where we define the boson functions
\begin{equation}
F_{Bn}(\beta \mu)=(1/\alpha) \int_{0}^{\infty} dx x^{n} N_{B}(\mu x/2) L,
\end{equation}
with
\begin{equation}
L =\ln|(1+x)/(1-x)|.
\end{equation}
The integrals (4.3) are not ultraviolet divergent and hence do not give
infinite $1/\varepsilon$ contribution. The fermion functions $F_{Fn}(a)$ are
defined by replacing in Eq. (4.3)
\begin{equation}
N_{B}(\mu x/2) \rightarrow N_{F}(\mu x/2).
\end{equation}

\vspace{5mm}
\subsection{Temperature dependent part of the coupling constant}

As pointed out in section 3.1 we encounter an ambiguity in determining the
renormalization constant $Z_B$ as a direct consequence of the lack of Lorentz
invariance. Here we write two formulas, one derived from
$\Pi_{00}$ and the other one from $\Pi_T$. From Eq. (3.15) we have  $(a=\beta
\mu)$
\begin{eqnarray}
&&-f^{(1)}(\mu,T)=(N/6+n_{F}/12)/a^{2}-2N[4F_{B0}(a)-F_{B2}(a)]\nonumber \\
&&-2n_{F}[F_{F0}(a)-F_{F2}(a)],
\end{eqnarray}
and the other one from Eq.(3.16) and $\Pi_{T}=(\Pi_{\mu \mu}-\Pi_{00})/2$
\begin{eqnarray}
&&-f^{(1)}(\mu,T)=(N/12+n_{F}/24)/a^{2}-N[7F_{B0}(a)+F_{B2}(a)]\nonumber \\
&&-n_{F}[F_{F0}(a)+F_{F2}(a)].
\end{eqnarray}
The result for Eq. (4.7) was reported in a short communication in [18](where an
overall factor of 1/2 was missing in the right-hand side of formulas (8) and
 (9)).

\vspace{5mm}
\section{Discussion}
\subsection{Review on the results of previous works}
\setcounter{equation}{0}
\renewcommand{\theequation}{5.\arabic{equation}}

Gendenshtein [5] calculated the QCD coupling constant at finite temperature in
the one-loop approximation by using the RG equation, the dimensional
regularization, and the covariant gauge with a space-like normalization
momentum
$p^\mu=(0,\mu)$ and obtained
\begin{equation}
A^{(1)}=(11N-2n_{F})/(3\alpha),
\end{equation}
as in Eq.\hspace{1mm}(3.31),
and
\begin{equation}
-f^{(1)}(\mu,T)=(N/3+n_{F}/6)/a^{2},
\end{equation}
where we have presented the results using our notations.

Kajantie et al. [6] studied the gauge field part of QCD with $N$ colors. They
used the $A_0=0$ gauge and defined two renormalization schemes by writing
\begin{eqnarray}
D_{\mu \nu}^{-1}=D_{0\mu \nu}^{-1}(Z_{A}-G/p^{2})-(F-G)P^{L}_{\mu \nu},
\end{eqnarray}
where $D_{\mu \nu}$ is the gluon propagator and $F, G$ and $P^{L}_{\mu \nu}$
come from the polarization tensor
\begin{equation}
\Pi_{\mu \nu}=FP^{L}_{\mu \nu}+GP^{T}_{\mu \nu}.
\end{equation}
In the ``magnetic prescription'' they fixed the propagator at the point
$p^{\mu}=(0,\mu)$ and had for the temperature depending function the expansions
(without quarks)
\begin{equation}
-f^{(1)}(\mu,T)=5N/(16a),\ \mbox{\rm for}\ a\ll 1,
\end{equation}
\begin{eqnarray}
-f^{(1)}(\mu,T)=N[17/(90a^{2})+83\alpha/(6300a^{4})],\ \mbox{\rm for}\ a\gg 1.
\end{eqnarray}
In the so-called ``electric prescription'' they derived the coupling constant
from $F$ as
\begin{eqnarray}
-f^{(1)}(\mu,T)=N[1/(3a^{2})-1/(4a)-22(\ln a)/(3\alpha)],\ \mbox{\rm for}\ a\ll
 1
\end{eqnarray}
\begin{eqnarray}
-f^{(1)}(\mu,T)=-N[1/(18a^{2})-11\alpha/(900a^{4})],\ \mbox{\rm for}\ a\gg 1.
\end{eqnarray}
Notice here the change of sign in the high $T$ behaviour. They concluded that
the "magnetic prescription" is more natural than the "electric" one because the
former uses the physical part of the gluon propagator. They also argued that
one has to choose $\mu\cong 3T$ in the thermal equilibrium. In that case the
effect of the temperature dependent parts becomes negligible.

Nakaggawa, Ni\'{e}gawa and Yokota [7] used the real-time formalism and studied
the scale-parameter ratios $\Lambda(a)/\Lambda$ derived from different vertices
in 4-flavour QCD. They used the covariant gauge and found that the ratios
derived from the three-gluon vertex and the gluon-ghost vertex show just the
opposite behaviour than the one derived from the gluon-quark vertex. Namely in
the former case one gets a growing ratio while in the latter case a decreasing
ratio.

Fujimoto and Yamada [8] used the real-time formalism and derived the
temperature
depending coupling constant from the gauge-independent Wilson loop. At $a\ll 1$
, it reads as (without quark contributions)
\begin{eqnarray}
-f^{(1)}(\mu,T)=C[1/(3a^{2})-1/(4a)-22(\ln a)/(3\alpha)].
\end{eqnarray}
The same authors have discussed the finite temperature RG equations [9] in the
one-loop approximation, using the real-time formalism, the covariant gauge and
the dimensional regularization. From the gluon propagator and three-gluon
vertex
they obtained:
\begin{eqnarray}
&&-f^{(1)}(\mu,T)=(C+T_{f})/(4a^{2})+C[-23/3F_{B0}(a)-3F_{B2}(a)-14/3G_{B0}(a)
\nonumber\\
&&+32G_{B2}(a)]+2T_{f}[F_{F0}(a)-3F_{F2}(a)+152/9G_{F0}(a)],
\end{eqnarray}
where $T_{f}=1/2$, and $C=N$ for $SU(N)$. The coefficient in front of $F_{B0}
(a)$
was originally 4/3. The error was pointed out in Ref. [10].

{}From the fermion propagator and fermion-gluon vertex:
\begin{eqnarray}
%% FOLLOWING LINE CANNOT BE BROKEN BEFORE 80 CHAR
&&-f^{(1)}(\mu,T)=(C+3C_{f}+2T_{f})/(12a^{2})-2(C_{f}+11C)F_{B0}(a)/3\nonumber\\
&&-16(C_{f}+10C)G_{B0}(a)/9-32CG_{B2}(a)+2(-5C/6+C_{f}/3+T_{f})\nonumber \\
&&\times F_{F0}(a)-4(8C_{f}+17C)G_{F0}(a)/9+16CG_{F2}(a),
\end{eqnarray}
where $C_{f}=(N^{2}-1)/(2N)$ for $SU(N)$.

{}From ghost propagator and ghost-gluon vertex:
\begin{eqnarray}
&&-f^{(1)}(\mu,T)=(C+T_{f})/(12a^{2})-C[7F_{B0}(a)+F_{B2}(a)\nonumber \\
&&+2G_{B0}(a)]-2T_{f}[F_{F0}(a)+F_{F2}(a)].
\end{eqnarray}
In the above equations the boson functions $F_{Bn}(a)$ are defined in
Eq.(4.3), while $G_{Bn}(a)$ are defined as
\begin{eqnarray}
G_{Bn}(a)=(2/\alpha)\int_{0}^{\infty} dx \int_{0}^{1} dy x^{n+1}N_{B}(\mu x/2)
/[x^2(y^2+3)-1].
\end{eqnarray}
Fermion functions $G_{Fn}(a)$ are defined by replacing $N_{B}(\mu x/2)$ by
$N_{F}(\mu x/2)$ in Eq. (5.13) as $F_{Bn}(a)$ was defined from Eq. (4.3). The
functions $G_{in}(A)(i=B,F)$ do not appear in our results Eqs. (4.6) and (4.7),
since they come from trigluon renormalization only. [Note an amusing
coincidence: $-f^{(1)}$ in Eq. (4.6)= $-f^{(1)}[G_{B0}(a)\rightarrow 0]$ of Eq.
(5.12), both of which are derived from very different prescriptions.]

Stephens et al. [19] performed a background field one-loop calculation of gauge
invariant beta functions at finite temperature, using the retarded/advanced
formalism developped by Aurenche and Becherrawy [28]. In terms of our
notations,
their result reads as:
\begin{eqnarray}
&&-f^{(1)}(\mu,T)=(N/12+n_{F}/24)/a^{2}-N[21/4F_{B0}(a)\nonumber \\
&&+F_{B2}(a)+7/2G_{B1}(a)]-n_{F}[F_{F0}(a)+F_{F2}(a)].
\end{eqnarray}
The numerical coefficients of the leading terms of the high-temperature
expansion and the fermion parts are in complete agreement with our result in
Eq.(4.7). In the boson parts, however, there are some numerical discrepancies
with our result.

\vspace{5mm}

\subsection{Comparison of asymptotic expansions}

In order to compare our results with those mentioned in section 5.1 we derive
the asymptotic expansions for $-f^{(1)}(\mu,T)$. The asymptotic expansions for
Eqs. (5.10)-(5.12) at $a=\mu/T\ll 1$ in the high-temperature regime read as
 follows:
\begin{eqnarray}
&&-f^{(1)}(\mu,T)=\pi C/(9\sqrt{3}a^{2})-23C/(48a)+[(-22\nonumber \\
&&+\pi/3\sqrt{3})C+(18-38\pi/9\sqrt{3})T_{f}](\ln a)/(3\alpha),
\end{eqnarray}
\begin{eqnarray}
&&-f^{(1)}(\mu,T)=[(1-\pi/\sqrt{3})C+3C_{f}]/(12a^{2})-(11C+C_{f})/(24a)
\nonumber\\
&&-[(23+82\pi/9\sqrt{3})C+(4-8\pi/9\sqrt{3})C_{f}+4T_{f}](\ln a)/(3\alpha),
\end{eqnarray}
\begin{eqnarray}
-f^{(1)}(\mu,T)=-7C/(16a)-[C(22+\pi/\sqrt{3})-8T_{f}](\ln a)/(3\alpha).
\end{eqnarray}
Next we derive the asymptotic expansions for our results, i.e., Eqs. (4.6) and
(4.7) at $a\ll 1$, by using the following high temperature expansions [9]:
\begin{equation}
F_{B0}(a)=1/(16a)+(\ell_{1}-1)/\alpha,
\end{equation}
\begin{equation}
F_{B2}(a)=1/(12a^{2})+(\ell_{1}-1/6)/(3\alpha),
\end{equation}
\begin{equation}
F_{F0}(a)=-(\ell_{2}-1/2)/\alpha,
\end{equation}
\begin{equation}
F_{F2}(a)=1/(24a^{2})-(\ell_{2}-1/6)/(3\alpha),
\end{equation}
where
\begin{equation}
\ell_{1}=\ln (a/4\pi)+\gamma,
\end{equation}
\begin{equation}
\ell_{2}=\ln (a/\pi)+\gamma.
\end{equation}
We find the asymptotic formula from Eq.\hspace{1mm}(4.6)
\begin{eqnarray}
&&-f^{(1)}(\mu,T)=(N/3+n_{F}/6)/a^{2}-N/(2a)\nonumber \\
&&-[N(22\ell_{1}-71/3)-4n_{f}(\ell_{2}-2/3)]/(3\alpha).
\end{eqnarray}
The $T^{2}$ term coincides with the one of Gendenshtein [5], i.e.,Eq. (5.1).
The
gluon part in this asymptotic formula is consistent with the result of Elze et
 al. [17] derived in the background field method, which in our
notations reads
\begin{eqnarray}
-f^{(1)}(\mu,T)=N[1/(3a^{2})-1/(2a)-22(\ln a)/(3\alpha)+\dots],\ \mbox{\rm for}
\ a\ll 1.
\end{eqnarray}
This coincides with the result of Nadkharni [12], who has the terms up to
$O(1/a)$. The delicate reason why in the background field method we have a
factor -1/2 in front of the second term proportional to $1/a$, while in some
results [see Eqs. (5.7) and (5.9)] it is equal to -1/4 is clarified by Elze et
al. in [17].

Next from Eq.\hspace{1mm}(4.7) we find the asymptotic formula:
\begin{eqnarray}
%% FOLLOWING LINE CANNOT BE BROKEN BEFORE 80 CHAR
-f^{(1)}(\mu,T)=-7N/(16a)-[N(22\ell_{1}-127/6)-4n_{F}(\ell_{2}-5/12)]/(3\alpha).
\end{eqnarray}

The behaviour of the transverse polarization tensor $\Pi_{T}$ in the infrared
region is known to have a form [17]
\begin{eqnarray}
\lim \Pi_{T}(0,{\bf k})/\mu^{2}=g^{2}[-cN/a+O(\ln a)].
\end{eqnarray}
The factor $c$ has been calculated in different gauges. In the covariant
$\xi$-gauge its value is [27]
\begin{equation}
c=(9+2\xi+\xi^{2})/64,
\end{equation}
whereas in the temporal axial gauge it is [6]
\begin{equation}
c=5/16.
\end{equation}
Our formula (5.26) at small $a$ behaves as Eq.\hspace{1mm}(5.27) with
\begin{equation}
c=-7/16,\ \mbox{\rm for bosons},
\end{equation}
and
\begin{equation}
c=0,\ \mbox{\rm for fermions}.
\end{equation}
These results can be read also from the results of Refs.\hspace{1mm}[9] [see
Eq. (5.17)] and [12]. The relation (5.31) was also noticed by Elze et al.
 in [17]. We remark here that our $c$ is negative and hence no
spurious pole appears in the transversal propagator
\begin{eqnarray}
D_{Tij}(0,{\bf k})=-(\delta_{ij}-k_{i}k_{j}/{\bf k}^{2})/[{\bf k}^{2}-\Pi_{T}
(0,{\bf k})],
\end{eqnarray}
in contrast to the covariant and the temporal axial gauge cases.

It is a known fact that the $T^{2}$ terms are gauge independent and the gauge
parameter dependence starts at $\sim T$ order [13]. Thus the fact that the
$T^2$
terms in Eqs. (5.15) and (5.16) are different from others, e.g. Eq. (5.1)
should
not originate from the gauge choices.

\vspace{5mm}
\subsection{Concluding remarks}

In concluding the paper we give a few comments:

1) In sections 5.1 and 5.2 we have seen that the temperature dependence of the
QCD coupling constant is very sensitive to the prescription chosen. This is not
 a trivial issue, because all the results obtained hitherto also heavily
 depends on the vertex chosen (i.e.,the trigluon, the ghost-gluon, or the
 quark-gluon vertex) to satisfy the renormalization condition of the QCD
coupling constant. Furthermore in some gauges, e.g., in the Coulomb gauge, the
 transversality condition (3.3) on the polarization tensor does not hold and
 hence the structure of the polarization tensor in such gauges is different
 from that which satisfies the condition.

One of the reasons why we encounter different results in the literature is that
the broken Lorentz invariance has not been treated as we have done in Eqs.
(3.15) and (3.16). For example in deriving Eqs. (5.10)-(5.12) the authors of
 Ref.[9] have extracted the gluon and the vertex renormalization constants in
 front of Lorentz invariant structures not paying attention to the break down
of Lorentz invariance.

Another possible source for discrepancies is that the Ward identities are used
at finite temperature for different renormalization constants. Remember that
the derivation of the Ward identity is based on the gauge invariance and also
on the Lorentz invariance at $T=0$. Thus it is not a surprise that results
from different gauges or even (within the same gauge) from different vertices
are totally different.

The correspondence between the imaginary and real time formalisms has been
investigated in detail [14-16]. The choice of the formalism can also be the
 source of the discrepancies under study was pointed out in these references.

To clarify the issue of choosing a suitable renormalization prescription, one
would need to compute the two-loop contributions to the coupling constant at
finite temperature. It was shown, in particular, in the massive $O(N)$ scalar
model that the one-loop result is drastically changed by two-loop contributions
at high $T$ and in zero momentum limit [29].

2) As we have seen the function $f^{(1)}(\mu,T)$ shows a power-like
$T$-dependence instead of a logarithmic fall-off as a function of $\mu$. Thus
we have
\begin{eqnarray}
&&-[f^{(1)}(\mu,T)-f^{(1)}(\mu_{0},T_{0})]\nonumber \\
&&=\sum_{n=1,2} c_{n}[(T/\mu)^{n}-(T_{0}/\mu_{0})^{n}]+\dots.
\end{eqnarray}
However, if a relation $T=(const)\mu$ holds, where $(const) \simeq 1/3$ [6],
then one would have a logarithmic $T$-dependence like in a zero temperature
environment. Such a relation holds if one is dealing with the thermal
equilibrium. In a thermostat one cannot speak about individuals and thus $\mu$
is not a measurable quantity.

We have derived Eq.\hspace{1mm}(2.27) using two essentially different RG
equations (2.23) and (2.24). Suppose we had used only one RG equation for $T$,
i.e., Eq. (2.24), then we would have instead of Eq. (5.33)
\begin{eqnarray}
&&-[f^{(1)}(\mu,T)-f^{(1)}(\mu_{0},T)]\nonumber \\
&&=\sum_{n=1,2} c_{n}[(T/\mu)^{n}-(T/\mu_{0})^{n}]+\dots.
\end{eqnarray}
In this case we would have a power-like dependence even with the relation
$T=(const)\mu$.

3) Since in a thermostat $T$ is a measurable quantity, though $\mu$ is not as,
stated in 2), we should be able to define the coupling constant as a function
 of only $T$. This could be done by calculating an expectation value of the
 coupling constant. In that case as a probability density function we could
 use either the Bose or the Fermi distribution. However such an expectation
 value of the coupling constant could not be used to a scattering process in a
thermostat wherein a particle with some definite energy enters.

We would like also to mention that all the machinary for the evolution of the
running coupling constant at finite temperature can be analogously applied for
the case of a quantum field theory(such as QCD) at finite energy as formulated
in [30].

In conclusion, we note in accordance with the previous observations that there
is in fact no unique way to define a temperature depending QCD coupling
constant
and the issue of finding its reasonable prescription is left as a subject of
 further investigation.

\newpage
\noindent{\Large Appendix \\}
\setcounter{equation}{0}
\renewcommand{\theequation}{A.\arabic{equation}}
\vspace{5mm}

The formulas for the integrals which we encounter in the text are assembled in
this Appendix.
\begin{equation}
\int d^{4}p/(2\pi)^{4}[1/p^{2}]|_{vac}=0,
\end{equation}
\begin{equation}
\int d^{4}p/(2\pi)^{4}[1/p^{2}]|_{boson}=iT^{2}/12,
\end{equation}
\begin{equation}
\int d^{4}p/(2\pi)^{4}[1/p^{2}]|_{fermion}=-iT^{2}/24,
\end{equation}
with $\mu_{ch}=0$.

In the following we use the static space-like prescription specified by
$k_{0}=0$ and $k^2=-\mu^2$:
\begin{equation}
\int d^{4}p/(2\pi)^{4}(1/D)|_{boson} =-2iF_{B0}(a),
\end{equation}
where $F_{B0}(a)$ is defined in Eq.\hspace{1mm}(4.3).
\begin{equation}
\int d^{4}p/(2\pi)^{4}[(kp)/D]|_{boson}=-i\mu^{2}F_{B0}(a),
\end{equation}
\begin{eqnarray}
\int d^{4}p/(2\pi)^{4}(p_{0}^{2}/D)|_{boson} =-i(\mu^{2}/2)F_{B0}(a).
\end{eqnarray}

For the fermions we have similar results as in Eqs.\hspace{1mm}(A.4), (A.5)
and (A.6) in which we just replace $N_B(\mu x/2)$ by $N_F(\mu x/2)$ and change
the sign in front of the equations.

\newpage
\begin{center}
\large {\bf {REFERENCES}}
\end{center}
\begin{enumerate}
\item L. Dolan and R. Jackiw, Phys. Rev. {\bf D9}, 3320 (1974).
\item S. Weinberg, Phys. Rev. {\bf D9}, 3357 (1974).

\item     For a review: N.P.Landsman and Ch.G. van  Weert, Phys. Rep.
{\bf 145},141 (1987).
\item     J.Kapusta, {\em Field Theory at Finite Temperature and Density},
          (Cambridge University Press, Cambridge, 1988).
\item     L.E.Gendenshtein, Sov. J. Nucl. Phys. {\bf 29}, 841 (1979).
\item     K.Kajantie and J.Kapusta, Ann. Phys. {\bf 160}, 477 (1985);K.Enqvist
          and K.Kajantie, Mod. Phys. Lett. {\bf A2}, 479 (1987).
\item     H.Nakkagawa and A.Ni\'{e}gawa, Phys. Lett. {\bf B193}, 263 (1987);
          {\bf B196}, 571(E) (1987); H.Nakkagawa, A.Ni\'{e}gawa and H.Yokota,
          Phys. Rev. {\bf D38}, 2566 (1988); H.Nakkagawa, H.Yokota and
A.Ni\'{e}gawa, Phys. Rev. {\bf D38}, 3211 (1988); H. Nakkagawa, A. Ni\'egawa
and H. Yokota, Phys. Lett. {\bf B244}, 63 (1990).
\item     Y.Fujimoto and H.Yamada, Phys. Lett. {\bf B212}, 77 (1988).
\item     Y.Fujimoto and H.Yamada, Phys. Lett. {\bf B200}, 167 (1988);
          Phys. Lett. {\bf B195}, 231 (1987).
\item     R.Baier, B.Pire and D.Sciff, Phys. Lett. {\bf B238}, 367 (1990).
\item     K.Enqvist and K.Kainulainen, Z. Phys. {\bf C53},87 (1992).
\item     S.Nadkharni, Phys. Lett. {\bf B232}, 362 (1989).
\item     Y.Fujimoto and H.Yamada, Z. Phys. {\bf C40}, 365 (1988).
\item     R.Kobes, Phys. Rev. {\bf D42}, 562 (1990).
\item     R.Baier, B.Pire and D.Schiff, Z. Phys. {\bf C51}, 581 (1991).
\item     H.Nakkagawa, A.Ni\'{e}gawa and H.Yokota, Phys. Lett. {\bf B244}, 63
          (1990).
\item     H.Elze, U.Heinz, K.Kajantie and T.Toimela, Z. Phys. {\bf C37}, 305
          (1988); U.Heinz, K.Kajantie and T.Toimela, Ann. Phys. {\bf 176}, 218
          (1986); H.Elze, K.Kajantie and T.Toimela, Z. Phys. {\bf C37}, 601
          (1988).
\item     J.Antikainen, M.Chaichian, N.R.Pantoja and J.J.Salazar, Phys. Lett.
          {\bf B242}, 412 (1990).
\item     M.A.van Eijck, C.R.Stephens and Ch.G.van Weert, Mod. Phys. Lett.
          {\bf A9}, 309 (1994);D.O'Connor, C.R.Stephens and F.Freire, Mod.
Phys.
          Lett. {\bf A8}, 1779 (1993).

\item     B.S.DeWitt, Phys. Rev. {\bf 162}, 1195 (1967).
\item     L.F.Abbott, Nucl. Phys. {\bf B185}, 189 (1981).
\item     F.J.Yndurain, {\em Quantum Chromodynamics} (Springer-Verlag, Berlin,
          1983).
\item     P.Pascual and R.Tarrach, {\em QCD: Renormalization for the
          Practitioner}\\
          (Springer-Verlag, Berlin, 1984).

\item     G.'tHooft and M.Veltman, Nucl. Phys. {\bf B44}, 189 (1972).
\item     C.W.Bernard, Phys. Rev. {\bf D9}, 3312 (1974).
\item     K.Kinslinger and M.Morley, Phys. Rev. {\bf D13}, 2771 (1976).
\item     O.K.Kalashnikov and V.V.Klimov, Sov. J. Nucl. Phys. {\bf 31}, 699
          (1980); Sov. J. Nucl. Phys. {\bf 33}, 443 (1981).
\item     P.Aurenche and T.Becherrawy, Nucl. Phys. {\bf B379}, 259 (1992).
\item     K.Funakubo and M.Sakamoto, Phys. Lett. {\bf B186}, 205 (1987).
\item     M.Chaichian and I.Senda, Nucl.Phys. {\bf B396}, 737 (1993);
          M.Chaichian, H.Satz and I.Senda, Phys. Rev. {\bf D49}, 1566 (1994).

 \end{enumerate}

\newpage
\begin{center}
\large {\bf {Figure Caption}}
\end{center}
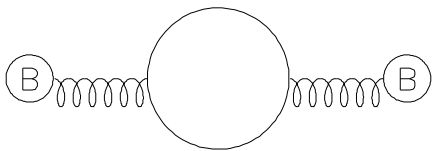
   Fig.1. Diagrams for the one-loop calculation of the background field
 renormalization factor $Z_{B}$: a) Gluon loop; b) Ghost loop; c) Gluon loop
 from the 4-gluon vertex; d) Ghost loop from the 2-gluon 2-ghost vertex;
e) Fermion loop. Wavy lines are quantum gauge field propagators.
 Wavy lines terminating in a "B" represent external gauge particles. Solid
 lines are fermion propagators and dashed lines represent ghost propagators.

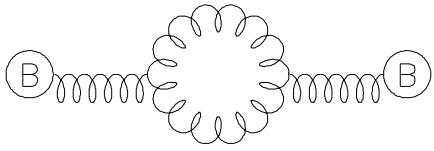
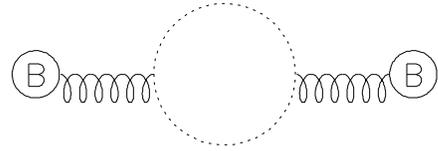
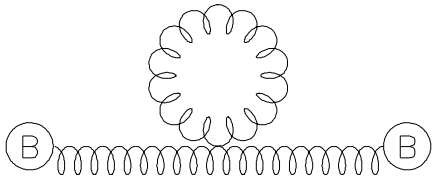
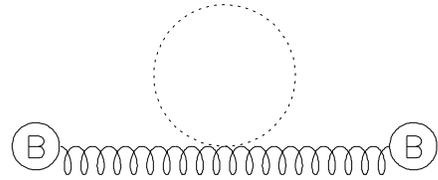
\end{document}